# Acceptable risks in Europe's proposed AI Act

Reasonableness and other principles for deciding how much risk management is enough

Henry Fraser[1] and José-Miguel Bello y Villarino[2]

*This paper critically evaluates the European Commission's proposed AI Act's approach to risk management and risk acceptability for high-risk AI systems that pose risks to fundamental rights and safety. The Act aims to promote "trustworthy" AI with a proportionate regulatory burden. Its provisions on risk acceptability require residual risks from high-risk systems to be reduced or eliminated "as far as possible" (AFAP), having regard to the "state of the art". This criterion, especially if interpreted narrowly, is unworkable and promotes neither proportionate regulatory burden, nor trustworthiness. By contrast the Parliament's most recent draft amendments to the risk management provisions introduce "reasonableness", cost-benefit analysis, and are more transparent about the value-laden and contextual nature of risk acceptability judgements. This paper argues that the Parliament's approach is more workable, and better balances the goals of proportionality and trustworthiness. It explains what reasonableness in risk acceptability judgments would entail, drawing on principles from negligence law and European medical devices regulation. And it contends that the approach to risk acceptability judgments need a firm foundation of civic legitimacy: including detailed guidance or involvement from regulators, and meaningful input from affected stakeholders*

## I. Introduction

The European Commission's Proposal for a Regulation Laying Down Harmonised Rules on Artificial Intelligence, issued in April 2021, is the most substantial venture towards AI regulation in the world.[3] The Proposal positions itself as a "risk-based" framework for "trustworthy AI", promising:

> a balanced and proportionate horizontal regulatory approach to AI that is limited to the minimum necessary requirements to address the risks and problems linked to AI, without unduly constraining or hindering technological development or otherwise disproportionately

---

[1] Dr Henry Fraser is a research fellow at the Queensland University of Technology node of the Australian Research Council Centre of Excellence for Automated Decision-Making and Society (ADM+S).
[2] Dr José-Miguel Bello y Villarino is a research fellow at the University of Sydney node of ADM+S, and a member of the Diplomatic Corps of Spain (on leave).
[3] *Proposal for a Regulation of the European Parliament and of the Council Laying down Harmonised Rules on Artificial Intelligence (Artificial Intelligence Act) and Amending Certain Union Legislative Acts*, COM/2021/206 final, 21 April 2021.

increasing the cost of placing AI solutions on the market.[4]

This paper critically evaluates the approach to risk management, and in particular risk acceptability judgements, for "high-risk" AI systems in article 9 of the Proposal. It compares the Commission's original text ("the Commission's Proposal") to amendments proposed by the Council of the European Union (the "Council's amendments") and European Parliament (the "Parliament's amendments") – which will form the basis of tripartite negotiations on the final text.[5] It argues that the approach to risk acceptability in the Commission's Proposal is unworkable. And it explains why amendments that would introduce reasonableness and cost-benefit considerations into risk acceptability judgements, as well as more stakeholder involvement and regulator guidance, should be adopted.

The approaches to risk acceptability in the Commission's Proposal on the one hand, and the Parliament's amendments, on the other, are quite different. Article 9(4) of the Commission's Proposal requires that risk management be such that "any" residual risks, as well as the overall residual risk from high-risk AI systems be judged "acceptable" (the article is set out in full in the appendix A to this paper). Residual risks are those risks that remain after the implementation of risk controls.[6] A key criterion for determining the appropriateness of risk management measures is that risks be reduced or eliminated "as far as possible" ("AFAP") through design and development. That is an exacting standard, and leaves no room for considerations of cost. By contrast, the Council's and especially the Parliament's amendments introduce "reasonableness", cost-benefit analysis and greater transparency about how risk acceptability judgements are made.[7]

This paper advocates for an approach in line with the Parliament's approach. In doing so, it makes three main arguments. Firstly, the Commission's requirement to reduce risks from high-risk AI systems "as far as possible" is neither workable nor consistent with the goals stated in the Explanatory Memorandum of the Commission's Proposal.[8] It creates too indeterminate a risk management obligation, which is inconsistent with the AI Act's commitment to proportionality of regulatory burden.[9] Perversely, that very uncertainty may encourage regulatees to fudge AFAP without principled justifications for where they draw the line. That is a poor formula for trustworthiness.

---

[4] Explanatory Memorandum, p.3.
[5] See, respectively, *Proposal for a Regulation of the European Parliament and of the Council Laying down Harmonised Rules on Artificial Intelligence (Artificial Intelligence Act) and Amending Certain Union Legislative Acts - General Approach* (Council of the European Union 2022) 2021/0106(COD) <https://data.consilium.europa.eu/doc/document/ST-14954-2022-INIT/en/pdf>; and *Amendments adopted by the European Parliament on 14 June 2023 on the proposal for a regulation of the European Parliament and of the Council on laying down harmonised rules on artificial intelligence (Artificial Intelligence Act) and amending certain Union legislative acts* s P9_TA(2023)0236 <https://www.europarl.europa.eu/doceo/document/TA-9-2023-0236_EN.html>
[6] ISO/IEC Guide 51:2014 Safety Aspects — Guidelines for Their Inclusion in Standards <https://www.iso.org/standard/53940.html> accessed 9 February 2023, 3.8.
[7] See above (n 5). The relevant amendments are Amendments 261 to 272 and are available at <https://www.europarl.europa.eu/doceo/document/TA-9-2023-0236_EN.html>

Secondly, amendments proposed to article 9 by the European Parliament, which introduce much clearer qualifications of reasonableness on risk management and acceptability, would make the regulatory burden of risk management more proportionate. Improving proportionality need not come at the cost of trustworthiness, because reasonableness calls for principled, transparent judgements about costs and benefits of risk management, whereas limits on risk management under AFAP are arbitrary.

Finally, since risk acceptability judgements in relation to high-risk AI systems are unavoidably value-laden, Europe needs to support them with an appropriate regulatory and civic architecture.[10] In order to promote truly trustworthy AI, regulators and stakeholders with the legitimacy and capability to weigh in on fundamental rights and the public interest must have input into risk acceptability judgements. There are positive developments in that direction, including in the Parliament's amendments, but more are needed.

At the time of writing, tripartite negotiations between the Commission, Council and Parliament on the final version of the AI Act have not yet commenced. Article 9(4) is still in that respect a moving target, although its main features and the issues for negotiation are now fairly clear. This paper accepts the risk of shooting at a moving target, given the opportunity to possibly influence the trilogue, and perhaps the Act's interpretation and implementation once finalised. In any event, the basic contributions of this article – critically evaluating different approaches to risk acceptability and explaining reasonableness in risk management - will remain relevant unless risk management provisions are totally removed. It may also be of interest to policymakers elsewhere as they consider whether to emulate

---

[8] See above (n 4).
[9] We will use "AI Act" or "Act" as a shorthand when discussing the overall concept of the regulation, its goals, and the final text (yet to be agreed at the time of writing), as distinct from the three separate drafts now under negotiation.
[10] On value-laden choices, see Michael Veale and Frederik Zuiderveen Borgesius, 'Demystifying the Draft EU Artificial Intelligence Act—Analysing the Good, the Bad, and the Unclear Elements of the Proposed Approach' (2021) 22 Computer Law Review International 97, 14.

Europe's approach. With all that in view, this paper proceeds as follows.

Part II zooms out from the arguments outlined above to provide background on AI, its risks and benefits, the basic mechanics of the proposed regulation, and to explain the importance of risk acceptability judgements given the regulatory architecture and goals.

Part III illustrates the problems with the approach to risk acceptability in the Commission's proposed text for article 9(4), particularly as regards the goal of proportionality. It describes the narrow interpretation the European Commission has given to the AFAP risk criterion in the context of medical device regulation in the past, and the dubious results of this interpretation.

Part IV sets out the key amendments to article 9 proposed by the Council and Parliament, observing that these changes use concepts of "reasonableness" to place limits on risk management. It then explores how reasonableness promotes proportionality of regulatory burden, drawing on jurisprudence from negligence, medical device regulation and beyond.

Part V considers the relative advantages and disadvantages of the Commission's and Parliament's approaches to risk acceptability. It explains why the Commission's AFAP risk criterion does not necessarily promote trustworthiness, despite its apparent stringency. It explains how a risk criterion informed by reasonableness (such as the approach in the Parliament's amendments) could still be consistent with a European, precautionary approach to risk: satisfying the AI Act's dual goals of proportionality and trustworthiness. And it briefly sketches the kinds of regulatory and civic architecture required to do so.

## II. How risk management and risk acceptability fit into the AI Act's goals and mechanics

Let us set the scene for the discussion in this paper. What is AI? What are its risks and benefits? How does the Commission's Proposal implement a "risk-based" approach? Where does article 9(4) fit in?

### 1. What is AI?

The Commission, Council and Parliament all define AI slightly differently, but all three definitions revolve around a common concept of a machine-based system capable of generating outputs that influence the environment, given explicit or implicit objectives.[11]

### 2. AI Benefits and risks

One of the fundamental motivations behind the AI Act, other than to promote a consistent approach to AI in Europe's internal market, is to realise AI's benefits while minimising its risks. This is evident not only in the Commission Proposal's explanatory memorandum, but also from the Act's origins in the European Commission's White Paper on Artificial Intelligence, published in February 2020.[12] The opening paragraph of the White Paper enumerates benefits and opportunities from AI across a wide range of domains – including increased productivity, better agriculture, and advances in medicine. It then immediately acknowledges that AI also "entails potential risks, such as opaque decision-making, gender-based or other kinds of discrimination, intrusion in our private lives or being used for criminal purposes."[13]

### 3. Risk-based approach

The Commission's Proposal responds to these risks (and opportunities) with a "risk-based" approach, intended to make regulatory burden proportionate to risk.[14] It does so by taking a tiered regulatory approach. Systems designated as unacceptably risky (e.g. manipulative systems) are prohibited.[15] Low risk systems attract only very light obligations of *transparency* for certain uses.[16] "High-risk AI systems" that pose risks to fundamental rights or safety are the meat in this regulatory sandwich, and their regulation occupies the bulk of the Proposal. They include systems capable of impacting access to education, jobs, government and other essential services, and systems used in law enforcement (among other applications).[17]

The precise contours of each tier of risk are different in the Parliament's and Council's amendments (e.g. the Parliament would expand the definition of "high risk AI systems" to encompass systems that pose 'significant' risks to health, safety, fundamental rights or the environment) but the tiered approach is common ground.[18] In all three texts, high-risk AI systems are subject to a set of risk control requirements set out in Title III of the Proposal, including data governance, documentation, oversight, accuracy, explanation and risk management.

---

[11] Commission's Proposal Art 3 and Annex III; Parliament's amendments (n 5), Amendment 165; Council's amendments, art 3.
[12] 'White Paper On Artificial Intelligence - A European Approach to Excellence and Trust' (European Commission 2020) COM (2020) 65 <https://ec.europa.eu/info/sites/info/files/commission-white-paper-artificial-intelligence-feb2020_en.pdf> accessed 3 July 2020.
[13] ibid.
[14] Julia Black and Robert Baldwin, 'Really Responsive Risk-Based Regulation' (2010) 32 Law & Policy 181.
[15] Title II.
[16] Title IV.
[17] Art 6 and Annex III.

## 4. Standards

Consistent with Europe's "New Legislative Framework", the Proposal contemplates that its requirements will be met through a "conformity assessment" against harmonised standards approved by European Standards Organisations.[19] The default will be self-certification, although provision may be made for certification by independent third parties.[20] Article 9(3) also provides that risk management "shall take into account the generally acknowledged state of the art, including as reflected in relevant harmonised standards or common specifications." Neither the Council nor Parliament's amendments depart significantly from the Commission's Proposal on this point. Standards bodies (primarily the European ones, namely CEN and CENELEC, but also global ones, such as ISO and IEC) will thus have significant influence over the day-to-day approach to risk acceptability.

## 5. The significance of the risk acceptability criterion

The amendments proposed by the Council and especially the Parliament are quite far-reaching, and introduce, among other things: specific provisions for general purpose AI and foundation models; fundamental rights impact assessments; broader definitions of high-risk AI systems which factor in environmental and health risk; and the creation of a dedicated AI Office.[21]

While these and other amendments will raise significant issues for negotiation between Commission, Council and Parliament, we focus here on risk acceptability because it is a crossroads where the AI Act's goals of proportionality and trustworthiness come to a head. Article 9(4) is a normative statement about how much risk from (risky but useful) "high-risk" AI systems European society is prepared to bear, and how much risk management is enough. It is not just a standalone requirement. It bears on the degree of effort and expense required to comply with the other requirements for high-risk systems.[22]

---

[18] See e.g. Parliament's amendments (n 5), Amendment 234.
[19] See Art 40; Art 65(6)(b). Regarding the NLF, now under review, see European Commission, 'New Legislative Framework' (*Internal Market, Industry, Entrepreneurship and SMEs - European Commission*, 2021) <https://ec.europa.eu/growth/single-market/goods/new-legislative-framework_en> accessed 13 August 2021.
[20] See e.g. Commission's Proposal, art 33.
[21] See e.g. Parliament's amendments (n 5), Amendment 4, Amendment 92, Amendment 122; Council's amendments, p 4.

Despite the pivotal importance of article 9(4), the provision has received surprisingly little scholarly attention. As far as we know, we were the first to grapple with it in detail, in a working paper in 2021.[23] Schuett, in this journal, has helpfully worked through how article 9 fits together as a whole, explained what each paragraph does, provided sensible working definitions for undefined terms, and suggested that risk management obligations be extended to encompass organizational risk management.[24] Mahler's work on risk management and proportionality in the Proposal is also relevant, although it does not address article 9(4) directly.[25] Because the Proposal defines AI so broadly, one of its key challenges, he contends, is to avoid a disproportionate regulatory burden, and to ensure coherence between several overlapping risk-oriented approaches.[26] Our work builds on this insight, zooming in on the need for proportionality in risk management.

## III. Problems with the Commission's 'AFAP' approach to risk acceptability – a historical perspective from medical devices regulation

The Commission's Proposal requires risks to be reduced "as far as possible" through risk management. What does that actually mean? We will try to answer that question in this part, describing and evaluating the Commission's very narrow historical approach to interpreting AFAP in the context of medical device regulation. We will also illustrate how the stringency of the criterion, perversely, creates uncertainty about how much risk management is enough and encourages stakeholders to fudge acceptability judgements.

---

[22] Art 8(2) of all three drafts (Commission, Council and Parliament) of provides that the risk management system in art 9 "shall be taken into account when ensuring compliance" with the other requirements in Title III.
[23] Henry L Fraser and Jose-Miguel Bello y Villarino, 'Where Residual Risks Reside: A Comparative Approach to Art 9(4) of the European Union's Proposed AI Regulation' [2021] Working Draft <https://papers.ssrn.com/abstract=3960461> accessed 7 December 2021.
[24] Jonas Schuett, 'Risk Management in the Artificial Intelligence Act' [2023] European Journal of Risk Regulation 1.
[25] Tobias Mahler, 'Between Risk Management and Proportionality: The Risk-Based Approach in the EU's Artificial Intelligence Act Proposal' [2020] Nordic Yearbook of Law and Informatics 247.
[26] Ibid.

Let us start with the history of AFAP. The language of article 9(4) in the Commission's Proposal, including the AFAP criterion, closely follows language from the General Requirements section of the European medical devices regulation.[27] However, it leaves out the General Requirements' definition of AFAP: "the requirement to reduce risks as far as possible means the reduction of risks *as far as possible without adversely affecting the benefit-risk ratio*."[28] References to the benefit-risk ratio are conspicuously absent from article 9. It is not clear why. The implication is that the standard is intended to be interpreted narrowly, excluding consideration of the opportunity costs of risk management.

The European Commission has also, in the past, explicitly rejected a broad interpretation of AFAP. It objected in 2010 to risk management provisions in EN ISO 14971, the medical device risk management standard, which treated AFAP as analogous to "as low as reasonably practicable" (ALARP), allowing cost-benefit considerations to enter into risk acceptability judgements (albeit with a heavy weighting toward risk reduction).[29] Subsequently, ISO, working together with the Commission, clarified that in the European context, all medical device risks had to be reduced AFAP "without there being room for economic considerations".[30]

The problems with a narrow approach to AFAP are evident from the reaction to the changes to the ISO standard. The industry group for "notified bodies" tasked with certifying medical devices against ISO 14971 reported that the change created uncertainty about "where to stop reducing risk before a product can be placed on the market" in Europe.[31] Other responses highlighted how unlikely businesses are to completely disregard "economic considerations."[32] Some risk management consultants simply advised their clients to conceal their cost-benefit analyses: "If you want to consider the economic impact of risk and keep track of business risks, you should create a separate document for that purpose."[33]

---

[27] Regulation (EU) 2017/745 of the European Parliament and of the Council of 5 April 2017 on medical devices, O.J. 2017, L 117/1, Annex 1.

[28] Medical devices regulation, ibid, para. 2. Our emphasis.

[29] ISO 14971: 2009, para 3.4 and D.8. At that time, the relevant regulatory instruments were the Medical Devices Directive (93/42/EEC), Active Implantable Medical Devices Directive (90/385/EEC) and the In Vitro Diagnostic Devices Directive (98/79/EEC). The AFAP risk criterion was already in place in the 'Essential Requirements' of these directives, which have now been superseded by the 'General Requirements' of the medical device regulations.

[30] Annex Z of EN ISO 14971:2012, content deviation 3. See also Peter Bowness, 'ISO 14971: 2012 - Risk Management' (British Standards Institute 2015) <https://www.bsigroup.com/globalassets/meddev/localfiles/it-it/webinars/bsi-md-iso-14971-risk-mgmt-webinar-presentation-25-march-2015.pdf> accessed 9 February 2023.

[31] 'Consensus Paper for the Interpretation and Application of Annexes Z in EN ISO 14971: 2012 v 1.1' (Notified Bodies Recommendation Group 2014) <https://www.team-nb.org/wp-content/uploads/2017/08/NBRG_WG-RM_Interim_NBmed_Consensus_Version_140812_1_1.pdf> accessed 22 December 2022.

The very impracticability of AFAP seems to encourage a lack of rigour in the approach to risk acceptability. In medical device risk management, demonstrating "state of the art" risk management measures is now taken as evidence in a conformity assessment of having reduced risks AFAP.[34] The same is likely to be the case for high-risk AI systems – given the explicit reference to "state of the art" in article 9(3).[35]

The meaning of "state of the art" is vague, however, to the point of being circular. The European Commission's *Draft standardization request in support of safe and trustworthy artificial intelligence* defines "state of the art" as follows:

> A developed stage of technical capability at a given time as regards products, processes and services based on the relevant consolidated findings of science, technology and experience. The state of the art does not necessarily imply the latest scientific research still in an experimental stage or with insufficient technological maturity. [36]

It is unclear whether standards are supposed to conform with the developed state of technical capability, or whether they themselves determine the state of the art. It is also not clear that the state of the art calls for the highest level of precaution feasible given current technology. Its meaning, at least in the medical device context, seems more akin to current good practice in industry.[37]

What is the import of all of this? If AFAP is interpreted in the AI context as it is for medical devices, it will pose several problems. Firstly, concerns about indeterminate risk-management costs are as relevant to AI systems as to medical devices. AI risks are 'emergent' and unpredictable and it is always possible to implement *just one more* measure to reduce risk.[38] Regulatees need more certainty about where to stop. Secondly, it is not clear that the "state of the art" or standards provide principled reasons for where to draw the line for risk management. Finally, the prospect of regulatees *pretending* to disregard economic considerations while *concealing* their cost-benefit analyses – in effect, fudging AFAP – is at odds with the Proposal's goal of trustworthy AI.

---

[32] 'AFAP As Far As Possible - Risk Reduction Requirement in Medical Devices' (*PresentationEze*) <https://www.presentationeze.com/blog/afap-as-far-as-possible-risk-reduction/> accessed 6 April 2022.
[33] BONEZONE Editors, 'How to Approach Risk Management Under ISO and MDR Updates' (*BONEZONE*, 5 March 2020) <https://bonezonepub.com/2020/03/05/how-to-approach-risk-management-under-iso-and-mdr-updates/> accessed 20 December 2022.
[34] We are indebted to John Lafferty of SQT Training Ltd and Northridge Quality & Validation for his help in understanding the approach to the "state of the art".
[35] See above part II, 5.
[36] *Draft standardisation request to the European Standardisation Organisations in support of safe and trustworthy artificial intelligence* (European Commission, December 2022), Annex II, note 1.
[37] Raje, 'Medical Device Risk Management for the Change from ALARP to AFAP' (*StarFish Medical*, 19 December 2013) <https://starfishmedical.com/blog/medical-device-risk-management-and-the-change-from-alarp-to-afap/> accessed 6 April 2022.

## IV. 'Reasonableness' in Parliament's proposals regarding risk management and risk acceptability

Changes proposed by both Council and Parliament to article 9 indicate that they are alive to the problems of indeterminacy of risk management and risk acceptability in the Commission's proposal. Both the Council and (to a much greater extent) the Parliament use the language of reasonableness to keep the burden of risk management in proportion to risk. In this part, we set out the key amendments proposed by each that bear on article 9(4), before exploring jurisprudence from negligence and medical devices regulation that gives a sense of how reasonableness brings risk-benefit and cost-benefit analysis into risk acceptability judgements.

The overarching obligation to identify, analyse and manage risks comes from art 9(2) of the Commission's Proposal. The Council's amendments limit the application of art 9(2) (and consequently the whole of article 9) to only those risks "most likely to occur" and state that risk management should "concern only those [risks] which may be *reasonably* mitigated or eliminated through the development or design of the high-risk AI system, or the provision of adequate technical information". [39]

The Parliament's amendments place even more emphasis on reasonableness and proportionality in risk management, and engage more directly with article 9(4) along with the rest of article 9.[40] Where the Council proposes to limit risk management only to risks "most likely to occur" (perhaps too narrow a scope), the Parliament proposes more balanced scope limitations for risk management. It envisages that the obligation to identify, estimate and evaluate risks apply to "reasonably foreseeable risks" (Amendment 263), and the obligation to evaluate risks emerging from post-market monitoring apply only to "significant" risks (Amendment 265).

---

[38] Andrew D Selbst, 'Negligence and AIs' Human Users' (2020) 100 Boston University Law Review 1315.
[39] Council's amendments, (n 3).
[40] Parliament's amendments (n 3).

In addition to limiting the scope of risk management, the Parliament also proposes to limit its burden. A proposed change to article 9(2)(d) qualifies the requirement to adopt risk management measures so that only "appropriate and targeted" risk management measures are required (Amendment 266). An amendment to article 9(3) adds wording to the effect that when regulatees consider the effects of risk management, they should have regard to mitigating risks effectively while "ensuring an appropriate and proportionate implementation of the requirements" (Amendment 267). Finally, proposed text in article 9(5) indicates that risk management measures should be weighed against the potential benefits and intended goals of the system (Amendment 273).

Parliament's changes to article 9(4) indicate a similar intent. Rather than requiring risk management to be such that risks are judged "acceptable" and reduced AFAP, the proposed wording is "*reasonably* judged to be acceptable" and "elimination or reduction of identified risks *as far as technically feasible* through adequate design and development of the high-risk AI system, involving when relevant, experts and external stakeholders" (Amendment 269).

What emerges from all of this is a contest between a narrow risk acceptability criterion – the Commission's AFAP - and more flexible approaches proposed by the Council and especially the Parliament, which take into account the significance of risks, the burden of risk management, and the value-laden nature of risk acceptability judgements. There will be many matters for the Parliament, Council and Commission to resolve in the trilogue, but the scope and burden of risk management and risk acceptability will be an important one. An understanding of reasonableness is relevant to the negotiations on the final form of article 9 and 9(4) and, ultimately, its implementation. It is important to unpack, then, what reasonableness entails.

Reasonableness permits risks to be judged in context, rather than according to a rigid, abstract standard. A possible pan-European approach to "reasonableness" in tort law is summarised in article 4:102(1) of the *Principles of European Tort Law* (**PETL**):

> The required standard of conduct is that of the reasonable person in the circumstances, and depends, in particular, on the nature and value of the protected interest involved, the dangerousness of the activity, the expertise to be expected of a person carrying it on, the foreseeability of the damage, the relationship of proximity or special reliance between those involved, as well as the availability and the costs of precautionary or alternative methods. The essence of reasonableness is not vastly different as between European and common law doctrines, even if there are differences in emphasis.

In common law negligence, four factors come into play in assessing reasonableness: (i) likelihood of harm; (ii) severity of harm; (iii) cost of precautions; and (iv) impact of precautions on the utility of the activity.[41] The economic approach to tort law is also based on this four-factor approach.[42] The four-factor approach is less prevalent in civil law jurisdictions, but in France and Germany at least, the four factors are mentioned in the legal literature,[43] as well as in recent commentaries on AI and civil liability in Europe.[44] In practice, the balancing is rarely conducted with algebraic precision because of the difficulty of ascribing precise probabilities and values to possible outcomes.[45] Rather, the general objective is to avoid obvious disproportion between risk management and risk (in either direction).

Reasonableness supports proportionality in risk management because it provides a framework for principled, transparent weighing of the trade-offs involved.[46] Precautions against risk generally involve loss of money, time, convenience or value, including value to the public.[47] Applying reasonableness principles opens the door to consideration of cost-benefit ratio, risk-benefit ratio, and materiality of risk. That is to say, it permits (indeed encourages) judgements about whether the cost of a given risk management measure is worth the reduction in risk; whether risk management negatively impacts the overall benefit of an AI system; and whether risks are significant enough to warrant expenditure of finite risk management resources.

---

[41] *Bolton v Stone* [1951] 1 All ER 1078; *Morris v West Hartlepool Co. Ltd* [1956] AC 552 per Lord Reid; *United States v Carroll Towing Co.*, 159 F.2d 169, 173 (1947); Restatement (Third) of Torts: Physical and Emotional Harm, §3e.

[42] See Guido Calabresi, *The Cost of Accidents* (Yale University Press 1970); Richard A Posner, 'A Theory of Negligence' (1972) 1 The Journal of Legal Studies 29; Steven Shavell, *Economic Analysis of Accident Law* (Harvard University Press 2009).

[43] See Cees Van Dam, *European Tort Law* (Cees van Dam ed, Oxford University Press 2013) paras 805–1 <https://doi.org/10.1093/acprof:oso/9780199672264.003.0008> accessed 10 February 2023. See generally See also Christian von Bar, *The Common European Law of Torts*, Vol. 2 (Oxford: Oxford University Press, 2000), 225. Hein Kötz and Gerhard Wagner, *Deliktsrecht*, 11th edn. (Neuwied: Luchterhand, 2010, )59–74; Geneviève Viney and Patrice Jourdain, *Les Conditions de la Responsabilité*, 3rd edn. (Paris: Librairie Générale de Droit et de Jurisprudence, 2006), 477.

[44] Miriam Buiten, Alexandre de Streel and Martin Peitz, 'EU Liability Rules for the Age of Artificial Intelligence' (Centre on Regulation in Europe 2021); Henrique Sousa Antunes, 'Non-Contractual Liability Applicable to Artificial Intelligence: Towards a Corrective Reading of the European Intervention' (8 February 2023) <https://papers.ssrn.com/abstract=4351910> accessed 14 February 2023.

[45] Peter Cane and James Goudkamp, *Atiyah's Accidents, Compensation and the Law* (9th edn, Cambridge University Press 2018) <https://www.cambridge.org/core/books/atiyahs-accidents-compensation-and-the-law/fault-as-a-basis-of-liability/4453E444E682C557552405D90E4300E8> accessed 6 May 2022.

[46] W Kip Viscusi and Michael J Moore, 'Rationalizing the Relationship between Product Liability and Innovation', *Tort Law and the Public Interest* (1991) 109.

[47] Buiten, de Streel and Peitz (n 44) 39; Shavell (n 42).

Given the AI Act's overarching goal of proportionality, the reason to consider *cost-benefit ratio* is that increasing the cost of risk management for an activity or product increases the marginal cost of its production. That cost is likely to be passed on to the public.[48] ISO's guidance on the medical device risk management standard is consistent with negligence jurisprudence on this point:

> Economic practicability refers to the ability to reduce the risk without making the medical device an unsound economic proposition, because the risk control measures would make the medical device too expensive and therefore unavailable… **However, economic practicability should not be used as a rationale for the acceptance of unnecessary risk**.[49]

The passage shows that there is a difference between considering economic matters and using them to justify unnecessary risk. It also indicates economic considerations may have broader significance to matters of public interest (such as the availability of beneficial systems and devices).

AI systems with the greatest potential social value – improving medical triage, reducing bias in loan or employment or judicial decisions etc. – may pose the highest risks, because of the high-stakes domain where they are deployed. If an AI system would be beneficial to society (and especially if it would be beneficial to the enjoyment or protection of fundamental rights or safety relative to the status quo) it would be undesirable unduly to increase the cost of putting such a system into use.

The reason to consider *risk-benefit ratio* is that risk management may have utility costs, as well as economic costs.[50] The classic example of risk management reducing the utility of an activity is speed limits. Lower speed limits reduce the risk of car accidents, but also the overall usefulness to the public of cars and driving (speed, convenience, and the generation of all kinds of public good via rapid movement of people and resources). That is why we are not all driving around at 5 km per hour.

---

[48] Restatement (Third) of Torts: Physical and Emotional Harm, §3e.
[49] ISO TR 24972:2020 para C.4, our emphasis.
[50] Cane and Goudkamp (n 45) pt 2.4.5.

Similarly, current ISO guidance for medical device risk management (where AFAP still means AFAP without disturbing the risk-benefit ratio) comments on how risk management may reduce the utility of a medical device.[51] It explains, by way of example, that reducing the power of an electrosurgical unit below its effective level in order to reduce risks would have an overall negative effect on the balance between benefit and risk for the device. Because risk-benefit considerations are permitted for medical devices, such an intervention would not be required.[52]

According to the same logic, negative impacts of risk management on the usefulness of a high-risk AI system – e.g. decreases in its effectiveness in reducing errors in administrative, medical, or financial decision-making – should also factor into whether a risk has been reduced "as far as possible".

Risk-benefit ratio assessments may also militate in favour of precaution – meaning that systems might in some circumstances *not* be judged to be "reasonably" acceptable, even if risks are managed as well as they possibly can be. In the medical device context, the final acceptability judgement turns not only on whether risks have been reduced as far as possible, but also on whether the overall risk-benefit picture is acceptable. It is no good putting AI systems into use on the basis that their risks have been reduced "as far as possible" or "as far as technically feasible" if this still leaves an intolerably high level of risk relative to the benefits of the system.

Since risk management may both increase the expense (cost-benefit impact) and decrease the utility (risk-benefit impact) of useful activities, a third reasonableness consideration is *materiality or significance of risk*. As risks become less significant, risk management delivers diminishing returns.[53] This is presumably why risk management based on "risk scoring", where risks of low probability and severity did not require further reduction, was the norm for medical devices before the Commission's intervention in 2010.[54] It is also why negligence doctrine, at least in some jurisdictions, does not require

---

[51] ISO TR 24972:2020, para C.4.
[52] Ibid.
[53] Buiten, de Streel and Peitz (n 44).
[54] Vincent Crabtree, 'How to Handle Medical Device Risk Management and the Change from ALARP to AFAP' (*StarFish Medical*, 19 December 2013) <https://starfishmedical.com/blog/medical-device-risk-management-and-the-change-from-alarp-to-afap-old/> accessed 20 December 2022; 'AFAP As Far As Possible - Risk Reduction Requirement in Medical Devices' (n 32).

the implementation of any risk management measures for insignificant risks.[55]

The language of the proposed amendments to article 9 and 9(4), especially Parliament's, invites these kinds of cost-benefit, risk-benefit and materiality threshold analysis. The jurisprudence we have just described could helpfully inform a number of judgements required under the Parliament's amendments to art 9, including: what it means to "reasonably" judge a risk acceptable; judgements about what is "appropriate and proportionate" or "appropriate and targeted" risk management; the weighing of risk management measures against the benefits of a system; and the approach to setting the threshold of materiality for risk management. In that respect, the Parliament's approach to risk management has the capacity to be not only more proportionate, but more principled, than the Commission's (as well as the Council's, which is more tentative in proposing pro-reasonableness amendments).

# V. Which approach to risk acceptability is better for trustworthiness?

So much for the question of which approach to risk acceptability best serves the goal of proportionate regulatory burden. What about the Act's other goal of trustworthiness? Here we consider the reasons why the Commission's approach to risk acceptability (AFAP), despite appearances, would not necessarily produce trustworthy outcomes; while the Parliament's approach, which is more explicit about cost-benefit judgements, and who is qualified to make them, is overall more likely to promote trust.

We start, however, with the objections that the Parliament's approach might have to overcome. Setting the threshold of acceptable risk is one of the most challenging and controversial tasks in risk governance.[56] Questions about acceptability of risks to human rights are not merely technical. They are contextual, social, normative, and political.[57] The disadvantage of reasonableness-style balancing is that it leaves room for disagreement about how to weigh different factors. It involves implicit or explicit judgements about the

---

[55] See e.g. *Civil Liability Act* 2002 (NSW), s 5B(1).
[56] Andreas Klinke and Ortwin Renn, 'The Coming of Age of Risk Governance' (2021) 41 Risk Analysis 544, 549.
[57] Fiona Haines, 'Regulation and Risk' [2017] Regulatory theory: Foundations and applications 181, 186; Veale and Borgesius (n 10).

value of the activity in question, and the interests affected by it.[58]

It is also conceivable that proportionality considerations might undermine trustworthiness. The European Parliament has said that citizens' trust in AI must be built on an ethics-by-design, "in line with the precautionary principle that guides Union legislation and should be at the heart of any regulatory framework for AI" and has called for the "utmost precaution" in relation to automated decisions in the areas of justice and law enforcement.[59] In essence, the precautionary principle favours erring on the side of caution, especially in the face of uncertain risks posed by new technologies.[60]

The European Commission has also noted that the application of the precautionary principle inevitably calls for political, normative judgement:

> Judging what is an "acceptable" level of risk for society is an eminently political responsibility. Decision-makers faced with an unacceptable risk, scientific uncertainty and public concerns have a duty to find answers.[61]

Risk acceptability judgements, which are qualitative in nature, rely heavily on skillful socio-cultural judgement.[62] They will not further the goal of trustworthiness unless those who make them have an imprimatur of legitimacy.

There is force in critiques of courts' abilities to judge difficult questions of social value and utility - particularly cost-benefit trade-offs involved in the design of complex systems.[63] For some, the fiction of the reasonable person is simply a way to disguise subjective judicial preferences.[64] Judges are expected to weigh uncertain values and trade off rights and interests, and courts have the constitutional legitimacy to do so – yet the weighing is still fraught.

How then can technologists implementing article 9(4) be expected to make such tricky value judgements, before a system has even been put into use? This is a strong objection to permitting reasonableness into the AI risk acceptability criterion (though we think it can be overcome). The objection is further fortified by the fact that commercial AI providers have an interest in resolving uncertainties in favour of lowering their costs.

---

[58] Cane and Goudkamp (n 45) pt 2.4.5; Van Dam (n 43) pt 8C. See generally David Rolph, Jason Varuhas and Penelope Crossley, *Balkin and Davis Law of Torts* (LexisNexis Butterworths 2021) pt 8.3, 8.29 <http://ebookcentral.proquest.com/lib/qut/detail.action?docID=6702645> accessed 31 May 2022.8.3, 8.29;
[59] European Parliament, Resolution of 20 Oct. 2020 with recommendations to the Commission on a framework of ethical aspects of artificial intelligence, robotics and related technologies, (2020/2012(INL), Art 3, Art 67.
[60] Jale Tosun, 'How the EU Handles Uncertain Risks: Understanding the Role of the Precautionary Principle' (2013) 20 Journal of European Public Policy 1517.
[61] Communication COM/2000/0001 final on the precautionary principle (European Commission, 2000) https://eur-lex.europa.eu/legal-content/EN/TXT/?uri=celex%3A52000DC0001 accessed 15 February 2023.
[62] Julia Black and Robert Baldwin, 'Really Responsive Risk-Based Regulation' (2010) 32 Law & Policy 181.
[63] See e.g. James A Henderson, 'Judicial Review of Manufacturers' Conscious Design Choices: The Limits of Adjudication' (1973) 73 Columbia Law Review 1531.
[64] Robyn Martin, 'A Feminist View of the Reasonable Man: An Alternative Approach to Liability in Negligence for Personal Injury' (1994) 23 Anglo-American Law Review 334.

One of the apparent advantages of the Commission's approach to risk acceptability, with the AFAP risk criterion, is that it appears to evade normative, political questions about the relative risks and benefits of any given high-risk system or set of risk management measures. Instead of making difficult value judgements, all that is required is to reduce the risk as far as possible (apparently without regard to cost).

AFAP also has symbolic force which may counteract commercial bias in risk management. The choice of the AFAP risk criterion signals that the European regulator will not tolerate cowboys. This might discourage regulatees from resolving uncertainties about costs and benefits in favour of selfish cost-saving. When it comes to the goal of trustworthy AI, this is a point in its favour.

Nonetheless, as we have argued above, risk management, even for high-risk systems, does have to stop somewhere. The AFAP criterion relies heavily on the state of the art and standards to mark that stopping point. The result is not to produce value-neutral risk management, but rather to cede control over important matters of public policy to technical standards bodies and the "state of the art".

If "state of the art" boils down to "best practice" or "good industry practice", then AI practitioners are likely to look to large technology companies, who have invested substantial resources into research on "ethical" or "fair, accountable and transparent" AI.[65] Yet, to take a recent example, Google just announced that it will "recalibrate" the level of AI risk it is willing to accept, and accelerate product reviews.[66] It is at least possible that Google's level of risk management was, as a matter of public interest, too high. The problem is that Google has openly stated that its motivation for recalibrating its risk practices is purely commercial: to ensure it is not outcompeted by OpenAI.[67] This is not a principled

---

[65] Meg Young, Michael Katell and PM Krafft, 'Confronting Power and Corporate Capture at the FAccT Conference | 2022 ACM Conference on Fairness, Accountability, and Transparency' [2022] 2022 ACM Conference on Fairness, Accountability, and Transparency (FAccT '22) <https://dl.acm.org/doi/10.1145/3531146.3533194> accessed 15 February 2023.
[66]  Nico Grant, 'Google Calls In Help From Larry Page and Sergey Brin for A.I. Fight', New York Times, Jan 20 2023 <https://www.nytimes.com/2023/01/20/technology/google-chatgpt-artificial-intelligence.html>
[67]   Ibid.

decisions about AI risk acceptability in the public interest.

It is also not clear that standards bodies have that political legitimacy or the expertise to make the value judgements involved in setting risk acceptability standards in relation to fundamental rights.[68] They are typically constituted by technology industry insiders, have little experience with human rights, and make minimal provision for participation by civil society and other stakeholders.[69] Standards generally govern the engineering of technical features and processes, rather than trade-offs between competing rights and interests in complex socio-political contexts.

If the Commission's risk acceptability criterion (AFAP x state of the art) is not likely to promote trustworthiness, is there a way of making the value-laden exercise contemplated by the Parliament more trustworthy? Two things are needed to make it so. Firstly, the approach to risk management and risk acceptability must be compatible with the precautionary principle. Secondly, European legislators and regulators have a duty to ensure that those who decide how risk acceptability for AI works have the capability and the civic, regulatory legitimacy to do so.

The first challenge can be met by stipulating with sufficient clarity that the costs of risk management and the quantum of risk should *not* be given equal weight in reasonableness assessments. Uncertainties must be resolved in favour of precaution. This message is conveyed to some extent by the choice of risk acceptability criterion in the Parliament's draft compromise amendments: "as far as technically feasible" – which has the advantage over AFAP of not having a documented history of literalistic interpretation by the Commission.

The message could be equally well (or better) conveyed, with the "as low as reasonably practicable" criterion that the European Commission rejected over ten years ago for medical devices.[70] ALARP is well understood by risk managers to require a precautionary approach. Generally, ALARP excuses risk managers only from risk management measures whose burden is *grossly disproportionate* to risk.[71] And (at least in some quarters) the approach to disproportion follows an exponential scale: the greater the risk, the higher the disproportion may be.[72] This is well suited to AI, which has a 'fat tail' of low probability, high gravity risks. ALARP also does not need to be read in light of other reasonableness requirements in article 9 in order to be practicable and proportionate (in contrast to the Parliament's current wording, AFATF, which does need that kind of 'reading down').

---

[68] Fiona Haines, *The Paradox of Regulation: What Regulation Can Achieve and What It Cannot* (Edward Elgar Publishing 2011) 149; Christine Galvagna, 'Discussion Paper: Inclusive AI Governance' (Ada Lovelace Institute 2023) <https://www.adalovelaceinstitute.org/report/inclusive-ai-governance/> accessed 15 May 2023. But see Kira JM Matus and Michael Veale, 'Certification Systems for Machine Learning: Lessons from Sustainability' (2022) 16 Regulation & Governance 177.
[69] Hadrien Pouget, 'The EU's AI Act Is Barreling Toward AI Standards That Do Not Exist' (*Lawfare*, 12 January 2023) <https://www.lawfareblog.com/eus-ai-act-barreling-toward-ai-standards-do-not-exist> accessed 13 February 2023.
[70] See above, part III.

The second challenge is one of regulatory capacity and civic legitimacy. The Proposal's architecture involves "regulated self-regulation", sometimes described as meta-regulation or co-regulation.[73] It is widely acknowledged in regulatory theory that de-centered regulation, a sharing of responsibilities between private and public entities (e.g. standards body and government regulators), is inevitable in a complex world.[74] But the allocation and structuring of regulatory discretion warrants careful consideration.[75] It should be attuned to the institutional and cultural characteristics and capabilities of the participants in regulation: and in this case there are doubts about whether standards bodies have the requisite capabilities.[76]

Two things would help here. Firstly, there needs to be input from affected groups and stakeholders on value-laden matters such as acceptability of risks to human rights.[77] Secondly, regulators with political and institutional legitimacy need to be more involved in making or at least guiding risk acceptability judgements.

---

[71] For the authoritative English explanation of ALARP and the 'grossly disproportionate' test, see *Edwards v. National Coal Board* [1949] 1 KB 704, 712; *Marshall v. Gotham Co Ltd* [1954] AC 360, 370 and 373.

[72] See e.g. UK Health and Safety Executive, "Principles and guidelines to assist HSE in its judgements that duty-holders have reduced risk as low as reasonably practicable" <https://www.hse.gov.uk/enforce/expert/alarp1.htm>accessed 6/7/23

[73] Peter Grabosky, 'Meta-Regulation' [2017] Regulatory theory: Foundations and applications 149; Haines (n 68); Ann Wardrop, 'Co-Regulation, Responsive Regulation and the Reform of Australia's Retail Electronic Payment Systems' (2014) 30 Law in Context 197.

[74] Black and Baldwin (n 62); Robert Baldwin, 'Better Regulation: The Search and the Struggle'; Wardrop (n 73); Neil Gunningham and Darren Sinclair, 'Smart Regulation' in Peter Drahos (ed), *Regulatory Theory: Foundations and applications* (ANU Press 2017); Christine Parker, 'Twenty Years of Responsive Regulation: An Appreciation and Appraisal' (2013) 7 Regulation & Governance 2.

[75] Rebecca Schmidt and Colin Scott, 'Regulatory Discretion: Structuring Power in the Era of Regulatory Capitalism' (2021) 41 Legal Studies 454, 461.

[76] Black and Baldwin (n 62); Baldwin (n 74); Wardrop (n 73); Gunningham and Sinclair (n 74); Parker (n 74).

[77] Karen Yeung, Andrew Howes and Ganna Pogrebna, 'AI Governance by Human Rights-Centred Design, Deliberation and Oversight: An End to Ethics Washing' in Marcus Dubber and Frank Pasquale (eds), *The Oxford Handbook of AI Ethics, Oxford University Press* (2019) 9; Nathalie A Smuha, 'Beyond a Human Rights-Based Approach to AI Governance: Promise, Pitfalls, Plea' (2021) 34 Philosophy & Technology 91.

At the time of writing, there are positive signs of both these two things – more provision for stakeholder involvement, and more input from regulators on both standards, and risk acceptability. The Commission's standardisation request to European Standards Organisations calls on them to take steps to ensure the involvement of small and medium enterprises and civil society organisations, and to ensure relevant expertise in the field of fundamental rights.[78] While this is promising, more could be done. Galvagna identifies prohibitive barriers to meaningful civil society participation in standardization. These include a lack of resources for engagement in standardization, restrictive eligibility requirements, lack of influence and status in relevant standards organisations, and lack of awareness and understanding of the standards process.[79] However, she recommends several changes, including amending the Regulation on European Standardisation to broaden the categories eligible for funding and mandated participation; funding for individuals to participate; and funding for a central hub to support civil society participation in standards.[80]

The Parliament's amendments seem to contemplate both stakeholder input on standards, and more guidance from regulators and stakeholders on risk acceptability. In the Parliament's version of article 9(4), risk acceptability judgements would involve "when relevant, experts and external stakeholders." Likewise, The Parliament's amendments explicitly provide for participation in standards making by a range of stakeholders (Amendments 103 and 104). They also make provision for the Commission to develop guidance on the requirements for high-risk systems, with the input of a dedicated AI Office (Amendment 529). Such a body would be well placed to provide guidance on article 9. Alternatively, there is the option of the European Commission issuing its own common specifications about risk management under article 41 of the Proposal, should standards fail to satisfy fundamental rights concerns.[81] Either guidance or common specifications would have an imprimatur of legitimacy and could articulate workable principles (ideally in line with the frameworks we have suggested here) to help both standards bodies and regulatees

---

[78] European Commission (2022) *AI Act: Draft Standardisation Request*, Annex II, accessed 14/4/23 <https://artificialintelligenceact.eu/wp-content/uploads/2022/12/AIA-%E2%80%93-COM-%E2%80%93-Draft-Standardisation-Request-5-December-2022.pdf>
[79] Galvagna (n 68).
[80] Ibid.
[81] Galvagna (n 68). See Parliament's amendments, Amendment 106.

decide how much risk management is enough.

## Conclusion

Commentators are justifiably concerned about under-regulating the activities of powerful, and sometimes reckless, AI developers.[82] One way in which the European Commission has signaled an intention to impose stringent requirements on high-risk AI, is with a forceful risk acceptability requirement in its AI Act Proposal: reduce risks from high-risk systems "as far as possible". The problem is, "as far as possible" does not mean what it says. In practice, the Commission's approach to risk management would allow a vaguely defined 'state of the art', untethered to principle, to dictate how much risk society should bear from high-risk AI systems.

Far better to recognize that contextual, qualitative, normative considerations of cost and benefit – and of what is reasonably practicable - must determine whether risks from a given high-risk AI system are acceptable in a given circumstance. Reasonableness has a rich jurisprudence that can assist in difficult judgements of risk acceptability. And there are existing risk acceptability criteria informed by reasonableness, such as "ALARP", that are already widely understood, and strike an appropriate, precautionary, balance between proportionality and trustworthiness.

As the Commission, Council and Parliament enter tripartite negotiations, they would do well to ensure that risk management obligations and the risk acceptability criterion are qualified by reference to reasonableness (as in the Parliament's draft language), rather than adopting the Commission's AFAP language. If there is room for compromise on the risk acceptability criterion, between the Commission's AFAP criterion and the Parliament's "as far as technically feasible", ALARP may be a "third way" worth exploring. Stakeholders and civil society should have meaningful input into what reasonable risk acceptability means in practice, and what standards say about risk acceptability. And a regulator with political legitimacy (such as the Parliament's proposed 'AI Office') should, in partnership with civil society, play a significant role in developing more detailed guidance on risk acceptability

---

[82] Yeung, Howes and Pogrebna (n 77); Julia Black and Andrew Douglas Murray, 'Regulating AI and Machine Learning: Setting the Regulatory Agenda' (2019) 10 European Journal of Law and Technology.

judgements.

# Acknowledgements

We are funded by Australian Research Council grant number CE200100005.

Our thanks go to Barry Wang and Rhyle Simcock for their research assistance. We are also grateful for the generous support and feedback of Prof Nic Suzor, Prof Kim Weatherall, Prof Henrique Sousa Antunes, Dr Raphaele Xenidis, Prof Mark Burdon, Prof Megan Richardson, Dr Alice Witt, Dr Anna Huggins, Dr Tiberio Caetano and Prof Karen Yeung.

The authors declare they have no competing interests.

# Appendix A

## Article 9(4)

The risk management measures referred to in paragraph 2, point (d) shall be such that any residual risk associated with each hazard as well as the overall residual risk of the high-risk AI systems is judged acceptable, provided that the high-risk AI system is used in accordance with its intended purpose or under conditions of reasonably foreseeable misuse. Those residual risks shall be communicated to the user. In identifying the most appropriate risk management measures, the following shall be ensured:

(a) elimination or reduction of risks as far as possible through adequate design and development;

(b) where appropriate, implementation of adequate mitigation and control measures in relation to risks that cannot be eliminated;

(c) provision of adequate information pursuant to Article 13, in particular as regards the risks referred to in paragraph 2, point (b) of this Article, and, where appropriate, training to users

In eliminating or reducing risks related to the use of the high-risk AI system, due consideration shall be given to the technical knowledge, experience, education, training to be expected by the user and the environment in which the system is intended to be used